\documentstyle[pre,aps,epsf,multicol]{revtex}
\draft
\begin{document}
\title{Direct Transition to Spatiotemporal Chaos in 
	Low Prandtl Number Fluids}
\author{Hao-wen Xi$^{1}$, Xiao-jun Li$^{2}$ 
	and J. D. Gunton$^{2}$ }
\address{$^1$Department of Physics and Astronomy, Bowling Green State
	University, Bowling Green, OH 43403.}
\address{$^2$Department of Physics, Lehigh University, 16 East Memorial Drive, 
	Bethlehem, PA 18015.}
\date{Revised \today}
\maketitle
\begin{abstract}
	We present a large scale numerical simulation of three-dimensional  
Rayleigh-B\'enard  convection near onset, under free-free
boundary conditions for a fluid of Prandtl number $\sigma=0.5$. 
We find that a spatiotemporally chaotic state emerges
immediately above onset, which we
investigate as a function of the reduced control parameter. 
We conclude that the transition from 
conduction to spatiotemporal chaos 
is second order and of ``mean field'' character. We also present a simple 
theory for the time-averaged convective current. Finally, we show that
the time-averaged structure factor satisfies a scaling behavior with respect
to the correlation length near onset.
\end{abstract}


\begin{multicols}{2}

	Pattern formation in non-equilibrium systems has become a major 
frontier area in science \cite{re:ch93,re:al95}. The richness
of this field has been significantly enhanced 
by the existence of
spatiotemporal chaos (STC) in various systems 
\cite{re:ch93,re:al95,re:exp,re:hu95,re:xgv,re:germ94,re:germ96}. 
STC is characterized by its extensive, irregular dynamics in both
space and time.  
It has been recognized in experiments
\cite{re:al95,re:exp,re:hu95} 
and numerical studies \cite{re:ch93,re:xgv,re:germ94} that a large aspect
ratio is  
essential for the occurrence of STC. 
Owing to the generic complication of its dynamics, 
theoretical understanding of STC relies heavily on some much simplified,
mathematical models of the real systems \cite{re:ch93}. 
Although much progress has been made, 
some fundamental concepts remain to be developed. 

A paradigm of 
pattern formation is 
Rayleigh-B\'enard convection (RBC) \cite{re:al95}, which occurs when
a thin horizontal fluid layer is heated from below.
In general, the dynamics of  RBC depends 
on the Rayleigh number $R$, the Prandtl number $\sigma$ of the fluid, 
and the aspect ratio (size/thickness)  $\Gamma$ of the system.
Busse and his collaborators \cite{re:bu78} 
have studied extensively the
stability domain of parallel rolls as a 
function of wavenumber $k$
and Rayleigh number $R$, for various $\sigma$.
It is well known that in a laterally infinite system with rigid-rigid 
boundaries, there exists a stable, time independent parallel roll state 
near the onset of convection for all $\sigma$.
In the case of {\it free-free} boundaries
at sufficiently low Prandtl numbers ($ \sigma  < $0.543),
Siggia and Zippelius \cite{re:si82} 
and Busse and Bolton \cite{re:bu84} found surprisingly that  parallel
rolls are unstable with respect to the skewed-varicose 
instability {\it immediately } above  onset.
Busse {\it et al.} \cite{re:bu92} further studied the possibility
of a direction transition from conduction to 
STC, but their aspect ratio ($\Gamma = 8$) is not large enough for a 
conclusive result.
Although the free-free boundaries are very difficult to control for detailed
experimental studies, one experiment with such boundary conditions  
has been reported  \cite{re:ggexp}.

In this paper, we present
the results of a large scale ($\Gamma = 60$) 
numerical simulation of the three
dimensional hydrodynamic equations, using the Boussinesq approximation,
for a low Prandtl number fluid ($\sigma$=0.5) with free-free boundary 
conditions. The same problem has been investigated 
before \cite{re:bu92,re:ar}, 
but the aspect ratio used by previous studies 
is too small for the occurrence of STC.
We find from our extensive numerical simulation 
that the convective state just above onset is spatiotemporally 
chaotic, which is evident from the snapshot images of the vertical 
velocity field and from the dynamical behavior of three important {\it global} 
quantities: the viscous dissipation energy, the thermal
dissipation energy and the work done
by the buoyancy force.  
This thus provides another
example of a direct transition to STC,
in addition to the K\"{u}ppers-Lortz transition \cite{re:hu95,re:kl69}, 
ac driven electroconvection and a few others \cite{re:germ96}.
Our method suggests that by studying the dynamical
behavior of some global quantities of systems which exhibit STC, one may
obtain valuable information about the 
temporal chaos of these systems. We also
measure the fractal dimensions of the global quantities and find a value of
about $1.4$. In addition, we
investigate the nature of the conduction to STC transition, as well as
certain properties of the spatiotemporally chaotic state.  
Our results for the
correlation length suggest that the
transition is second order, with a mean field power law behavior. 
In comparison recent experimental results for the 
K\"{u}ppers-Lortz transition are not consistent with a 
mean field power law behavior \cite{re:hu95}.
We present below a simple but rather accurate theory for the
behavior of the time-averaged convective current as a 
function of the reduced control
parameter.   Finally, we show
that the time-averaged structure factor (power spectrum) exhibits a
scaling behavior with respect to the correlation length similar to that found
in critical phenomena. 

The Boussinesq equations, which describe the evolution
of the velocity field $ \vec{u}(x,y,z,t)=(u,v,w)$  and the deviation of the
temperature field $\theta(x,y,z,t) $ from the conductive profile,
 can be written in dimensionless form as 
\begin{equation}
\left\{
\begin{array}{lll}
\nabla \cdot \vec{u} =0, \\
\partial \vec{u}/\partial t + (\vec{u} \cdot \nabla) \vec{u}=
-\nabla p + \sigma \theta \vec{e}_{z} + \sigma \nabla^{2} \vec{u}, \\
\partial \theta/\partial t + \vec{u} \cdot \nabla \theta
=\nabla^{2} \theta + w R,
\end{array}
\right.
\end{equation}
where $\vec{e}_{z}$ is the unit vector in the vertical z-direction.
In the idealized limit of a laterally infinite system, the
critical Rayleigh number 
$R_{c}=27 \pi^{4}/4$ and the onset wavenumber $k_{c}=\pi/\sqrt{2}$.
The efficient Marker-and-Cell (MAC) 
\cite{re:hw65,re:hc72} numerical technique is employed. 
(In developing our algorithm, we have carried out two tests: One is against
the theoretical results of Schl\"{u}ter {et al.} \cite{re:sc65}; the other is  
against the 
numerical results of Kirchartz and Oertel \cite{re:kir}. In both cases
agreements within $1\%$ have been found.)
The boundary conditions for the velocities $\vec{u}$ 
are free-slip at the upper and lower surfaces, and no-slip at the sidewalls.
The temperature deviation $\theta$ is zero on the top and bottom surfaces,
while the temperature gradient $\nabla \theta$ normal 
to the sidewall is set to zero. 
For the initial conditions of $\vec{u}$ and 
$\theta$, we have tried both random values
and values of Gaussian distribution, inside a possible range. 
Since no difference
has been found, the actual calculation is carried out with initial 
conditions of Gaussian distribution. 
Our parameters are $\sigma=0.5$ and $0.03 \le \epsilon \le 0.5$, 
where $\epsilon=(R-R_{c})/R_{c}$
is the reduced Rayleigh number. 
We use mesh points N$_{x} \times$ N$_{y} \times$ N$_{z}$
=256 $\times $ 256 $\times $18 and a grid size 
$\Delta x$=$\Delta y$=60/256, $\Delta z$=1/18 for an aspect
ratio $\Gamma$ =60 in the simulation. 
We have run for $360$
vertical diffusion times before collecting data. Considering that the normal
relaxation time to approach a steady state is about $10$ vertical diffusion
times [$\tau_{relax}= 2 (1+\sigma)/3\pi^{2} \sigma \epsilon$], 
we believe that we are well beyond any transient regime.

\narrowtext
\begin{figure}
\epsfxsize = 0.4 \textwidth
\epsfysize = 0.4 \textwidth
\makebox{\epsfbox{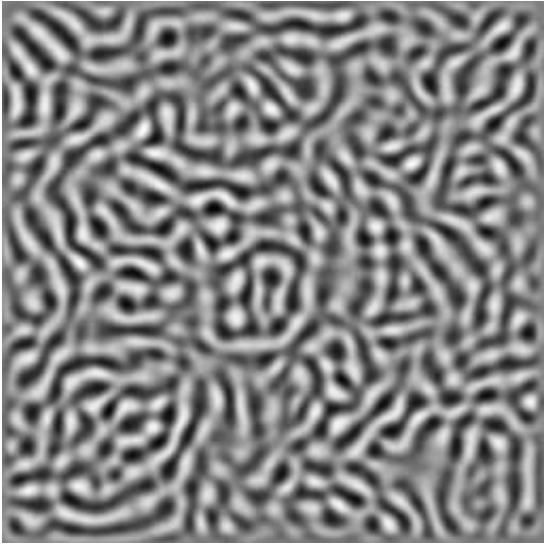}}
\medskip
\caption{A typical image of the spatially disorganized pattern
in the cell. Dark regions correspond to hot rising fluid and white
regions correspond to cold descending fluid. The 
vertical velocity field $w$(x,y,z=1/2) for $\epsilon=0.1$ is shown here.}
\end{figure}

For the low Prandtl number fluid studied here, the  
convective pattern near onset has an irregular space-time dependence. 
In Figure 1 we show a snap shot image of the vertical velocity 
field $w(x,y,z=1/2)$ from the numerical simulation at $\epsilon=0.1$. 
In this image, the apparently disorganized spatial pattern consists of  
superimposed rolls with many different orientations. The time evolution of 
these rolls is through an interface motion, which maintains the type of spatial
disorder shown in Figure 1.  Similar images are 
found for other values of $\epsilon$. It is obvious from such images that the
convection pattern near onset is random in space.

To illustrate the temporal chaos of the system, 
we now investigate the dynamics of the global quantities 
which characterize the underlying physics of Rayleigh-B\'enard
convection.  Using $\langle f\rangle $ 
to denote the average of $f$ over the whole system
and taking into account the boundary conditions as well as  
the incompressibility condition, we obtain $\frac{1}{2}
d\langle \vec{u} \cdot \vec{u}\rangle /dt=F_{2}(t)-F_{1}(t)$, and
$\frac{1}{2} d\langle \theta ^{2}\rangle /dt= F_{4}(t)-F_{3}(t)$, where
 (a) $F_{1}=\frac{1}{2} \sigma \langle (\partial u_{i}/\partial x_{j}+
\partial u_{j}/\partial x_{i})^{2}\rangle $ is the kinetic energy 
dissipated by the viscosity; (b) $F_{2}= \sigma \langle w \theta\rangle $ is 
the work done by the 
buoyancy force; 
(c) $F_{3}= \langle \nabla \theta \cdot \nabla \theta\rangle  $ is  the
dissipative thermal energy (generation of entropy) owing to 
temperature fluctuations; and 
(d) $F_{4}= R  \langle w \theta\rangle  = (R F_{2}/\sigma)$ is  
the flow of the entropy fluctuations carried  by 
the vertical velocity. It is clear  that in the special case of steady state
($\frac{\partial}{\partial t}=0$), one recovers the condition 
$F_{1}=F_{2}$ and $F_{4}=F_{3}$.  These global 
quantities provide us with a complete description of ``energy-balance'' in 
the Rayleigh-B\'enard system.

\begin{figure}[t]
\epsfxsize = 0.45 \textwidth
\epsfysize = 0.45 \textwidth
\epsfbox{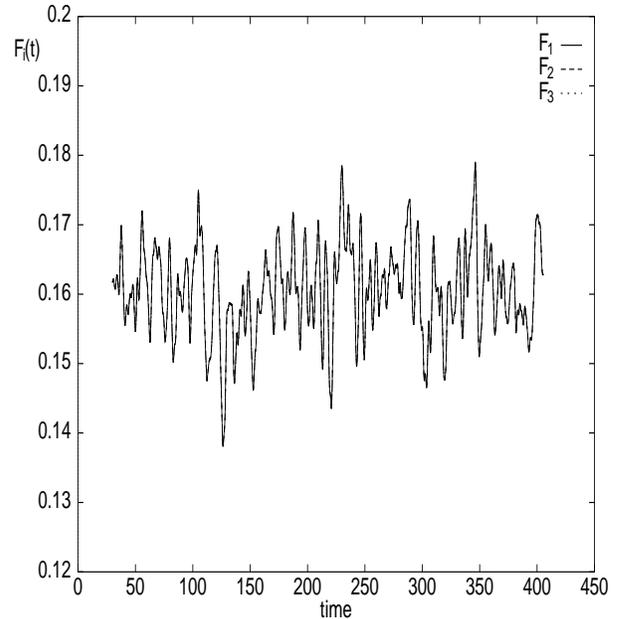}
\medskip
\caption{A plot of global 
quantities $F_{1}(t)$, $F_{2}(t)$ and $F_{3}(t)$ 
as functions of time for $\epsilon = 0.2$. 
Note that they lie on top of each other with differences, 
though substantial, too small to be seen. The  
time is in units of vertical thermal diffusion time $t_{v}=d^{2}/\kappa$
and the origin corresponds to $t = 330 t_v$ in real calculation.}
\end{figure} 

We plot a representative time series of these quantities, 
$F_{1}(t)$, $F_{2}(t)$ and $F_{3}(t)$, in Figure 2 for $\epsilon=0.2$. 
We have rescaled $F_{1}$, $F_{2}$ by $ \sigma R_{c}$,
and $F_{3}$ by $R R_{c}$ so that we have $F_{1}=F_{2}=F_{3}$ 
in a steady state. (Note that $F_{4}$ 
is simply related to $F_{2}$ by the factor R/$\sigma$.)  The
most important implication of this figure is the apparent chaotic
behavior of these quantities over the time interval that is
accessible to us. To be more concrete, we apply the Grassberger and Procaccia
 method \cite{re:gp83}
to compute the fractal dimensions $D_{f}$ for these quantities 
and find $D_f = 1.42 \pm 0.02$. 
This of course is different from the fractal dimension that
is normally used to characterize STC, which diverges with the system size.
We believe
that such global quantities might
provide a relatively simple way to characterize the temporally chaotic
nature of spatiotemporally chaotic states such as studied here, although
data over a longer time interval will be necessary for such an analysis.
It is also interesting to observe that the dynamics of 
these three quantities 
are {\it almost} exactly the same, i.e. 
$F_{1}(t) \simeq F_{2}(t)  \simeq F_{3}(t)$,
as shown in Figure 2. Although the differences among them are substantial
and beyond numerical uncertainties, 
they are too small to be seen under the scale
of Figure 2.
In fact, this is the case for all 
$\epsilon$ studied in the range $0.03 \le \epsilon \le 0.5$.
This is certainly a surprising result 
considering the irregular spatiotemporal state we observed. However,
a theoretical understanding of this
has been obtained, as outlined later.
This result also implies that the quantity  
$\Re = \sigma F_{3}/(2F_{2}-F_{1})$,
which is often used as a variational formulation to determine the critical
Rayleigh number, behaves as if the system is almost in a steady state. 

\begin{figure}
\epsfxsize = 0.45 \textwidth
\epsfysize = 0.45 \textwidth
        \makebox{\epsfbox{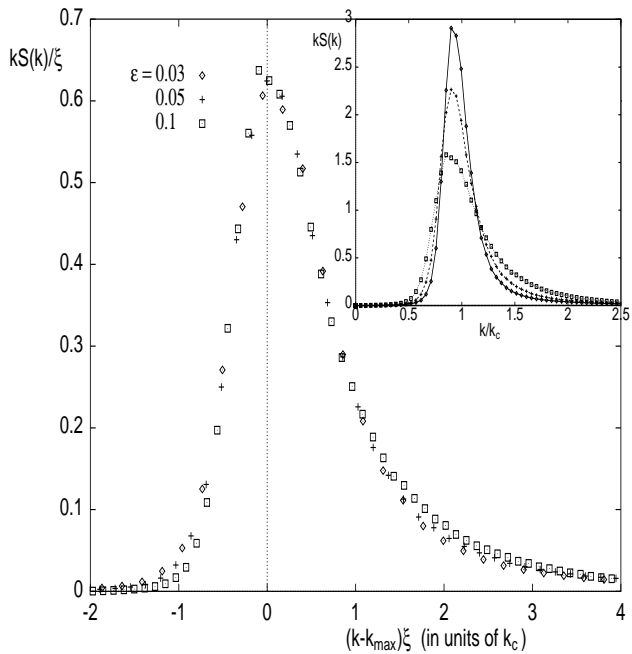}}
\vskip 0.3in
\caption{A plot of $k S(k)/\xi$ vs $x= (k - k_{max})\xi$ 
(in units of $k_c$), showing
scaling and the scaling function F(x) defined in the text. 
Insert: The time-averaged function $k S(k)$ vs. $k/k_c$  
for $\epsilon=0.03$, $0.05$ and $0.1$.}
\end{figure}

In order to gain more insight into the nature of the transition to STC
near onset, we have studied the two-dimensional structure factor (Fourier
 power spectrum). Since  
the snapshot  images of the patterns appears to be isotropic azimuthally, 
we calculate the azimuthally averaged structure factor, and 
then average the images over time, to obtain the time-averaged
structure factor $S(k)$.  The function $k S(k)$ 
is shown for several different values 
of $\epsilon$ in the insert in Figure 3. 
We also show in this figure that $k S(k)$ satisfies
a scaling behavior somewhat similar to that found in critical phenomena,
 namely,
$kS(k)/\xi=F[(k-k_{max})\xi]$, where $\xi$ is the correlation length
(defined below) and where we have normalized the integral
of $S(k)$ over $k$-space to be unity. Here $k_{max}$ is the wavenumber 
which maximizes $k S(k)$, 
which we choose to give the best fit to scaling.
Another interesting feature of the structure factor is associated with 
the power-law  behavior of S(k) $\sim k^{-3}$ for large wavenumber. 
This feature is observed over the range of $\epsilon$ studied here.
It is interesting to note that this is the same  
power-law behavior observed in phase separating systems, 
in two dimensions, where it is known as Porod's law. In both cases
it results from the linear behavior of the real space correlation function,
C(r), for small r, where this correlation function is the azimuthal average
of the  inverse Fourier transform of the structure factor.

\begin{figure}
\epsfxsize = 0.5 \textwidth
\epsfysize = 0.5 \textwidth
\epsfbox{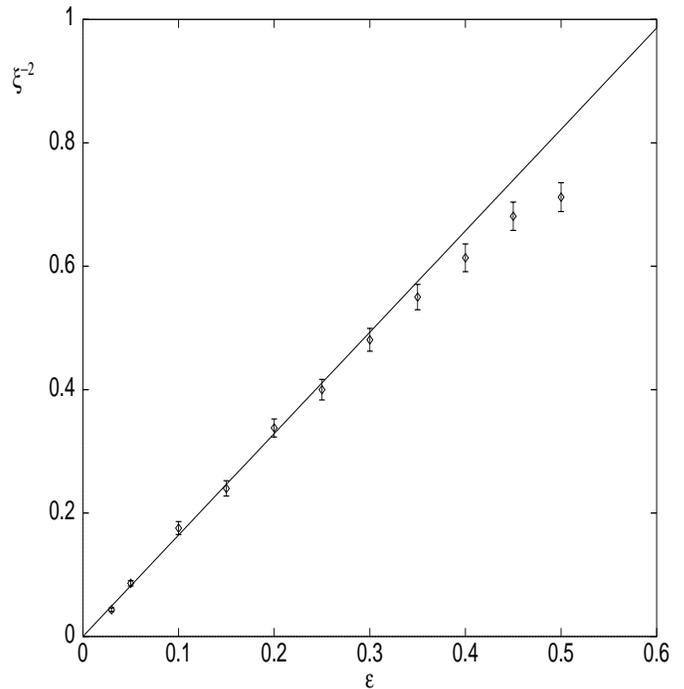}
\kern 10pt
\caption{A plot of $\xi^{-2}$ vs. $\epsilon$.  
The vertical bars indicate 
the standard deviation, and the solid line corresponds to 
$\xi^{-2}=  \xi_{0}^{-2} \epsilon$ with $\xi_{0}=0.78$. These data can also
be fitted with a nonconventional exponent: see the text for more detail.}
\end{figure}
 
We have also calculated the correlation length $\xi$ as a function
of the control parameter $\epsilon$, where we define the correlation length
through the variance of the wavenumber, i.e. 
$\xi=(\overline{k^{2}}-\overline{k}^{2})^{-1/2}$. 
The moment $\overline{k^{n}}$ is defined 
as $\overline{k^{n}}=\int |\vec{k}|^{n} S(\vec{k}) d^{2} \vec{k}/
\int S(\vec{k}) d^{2} \vec{k}$ and
$S(\vec{k})$ is the time-averaged structure factor.
We find that the correlation length $\xi$ appears to diverge as $\epsilon$
approaches the transition point, with a power-law behavior of 
$\xi= \xi_{0} (\epsilon-\epsilon_{c})^{-\nu}$ with $\nu =0.472 \pm 0.016$,
$\xi_{0}=0.82 \pm 0.04$ and $\epsilon_{c}=0.005$.
(The fact that $\epsilon_c$ is finite instead of zero is due to finite
size effects.)  
The behavior of the correlation length is also consistent with a mean field
power low exponent of $\nu =0.5$ and $\xi_{0}=0.78$. 
This is illustrated in Figure 4. The amplitude value 
$\xi_{0}$
is a factor of $3/2$ larger than the value 
$\xi_{0} = \sqrt{8/3 \pi^{2}}=0.52$ 
calculated from the curvature 
of the marginal stability curve.  

In order to investigate the heat transport in STC near onset, 
we calculate the Nusselt number $N(t)=1+\langle w \theta\rangle /R$, 
which describes the
ratio between the heat transport with and without convection, 
as a function of $\epsilon$. The quantity $N-1$ can be fit with a power
law behavior of the form $N-1=(\epsilon-\epsilon_{c})^{\mu}/\overline{g}$
with $\mu=1.034 \pm 0.025$, $\epsilon_{c}=0.012$ and
$\overline{g}=1.27 \pm 0.03$.  Figure 5 shows that the time-averaged 
$N-1$ is also consistent with a linear relation 
$(\epsilon-\epsilon_c)/\overline{g}$, where $\epsilon_c =0.01$ and
$\overline{g}=1.23$. (Again, owing to finite size effects, the value of
$\epsilon_c$ is nonzero.)  
We have also determined the time-averaged
vertical ``vortex energy'' $\Omega=\langle \omega_{z}^{2}\rangle $, 
where $\omega_{z}$ is
the vertical vorticity,
as a function of $\epsilon$ from our simulations and observed a 
power-law behavior of
$\Omega \sim \epsilon^{\lambda}$ with $\lambda = 5/2$.  

\begin{figure}
\epsfxsize = 0.45 \textwidth
\epsfysize = 0.45 \textwidth
\epsfbox{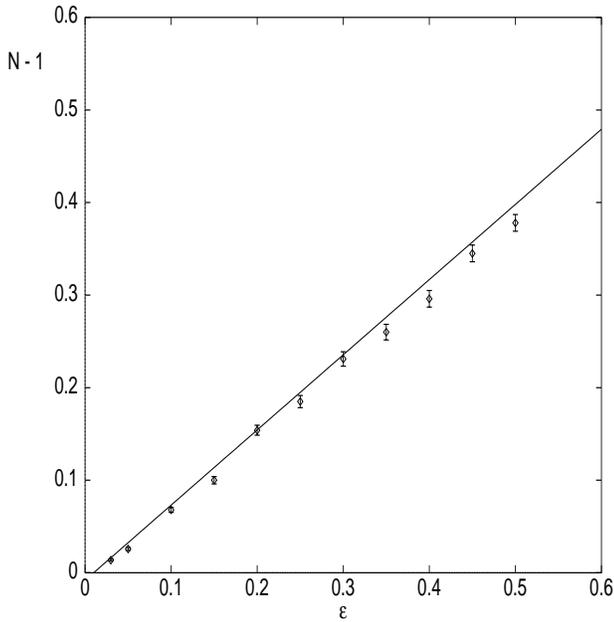}
\kern 10pt
\caption{Time-averaged $N-1$  vs $\epsilon$ near onset. 
The vertical bars indicate the standard deviation, and the solid line 
is the fit of $(\epsilon - \epsilon_c)/\overline{g}$ to the data 
with $\epsilon_c =0.01$ and 
$\overline{g}=1.23$. A slightly different fitting form is 
given in the text.}
\end{figure}

For theoretical understanding of some of our results, 
we notice that the velocity $\vec{u}$ and
the temperature derivation $\theta$ near onset can be approximated by an order
parameter $\psi(\vec{r})$ in two-dimensional space $\vec{r}$
multiplied by known
prefactors with $z$-dependence \cite{re:cr80}. It is then straightforward
to rewrite the global quantities $F_1$, $F_2$ and $F_3$, and the Nusselt
number $N$ in terms of $\psi(\vec{r})$. 
We further assume that only those modes inside the vicinity of $k_c$ are
excited and equally excited. 
From these approximations,  we confirm 
(after rescaling mentioned earlier) that $F_1(t) \simeq F_2(t) \simeq F_3(t)$.
We also obtain that 
\begin{eqnarray}
\overline{g}&=&g(-1) + (2/\pi)\int_{0}^{\pi} g(\cos \alpha) d \alpha
	\nonumber \\
	&=&0.855951+0.0458145 \sigma^{-1}+0.0709325 \sigma^{-2},
\end{eqnarray}
where $\alpha$ is the angle between $\vec{k}$ and some reference direction,
and we have used the explicit form of $g(\cos \alpha)$ given
by Schl\"{u}ter {\it et al.} \cite{re:sc65} for free-free
boundary conditions \cite{re:cr80}. For $\sigma$=0.5, 
we find $\overline{g}$= 1.2313,
which is in surprisingly good agreement with the numerical results. 
The theory of the vortex energy $\Omega \sim \epsilon^{5/2}$ is more 
complicated than the above. All this theoretical analysis will be
presented elsewhere. 

In summary, we have presented a large scale numerical simulation
of pattern formation in three dimensional Rayleigh-B\'enard convection.
We have calculated the spatial correlation length and the Nusselt number as
a function of the reduced control parameter, as well as the 
dynamics of the viscous energy,  
the dissipative thermal energy and the work done by the buoyancy force.
Our numerical studies suggest that the  transition from the
conduction state to spatiotemporal chaotic state near onset 
is a continuous (second order) transition, with mean field exponents
for the correlation length and the time-averaged convective current.
We have also demonstrated that 
the time-averaged structure factor satisfies a scaling behavior
with respect to the correlation length.  We believe that 
more studies of scaling in STC by systematic experiments 
for smaller $\epsilon$,
larger aspect ratio, and  with both free-free and rigid-rigid boundary
conditions will be challenging and valuable.

X.J.L. and J.D.G. are grateful to the National Science 
Foundation for support under Grant No. DMR-9596202. 
The computations were performed at the Pittsburgh 
Supercomputing Center and the Ohio Supercomputer Center.

\end{multicols}


\begin{thebibliography}{99}
\bibitem{re:ar}
W. Arter, J. Fluid Mech. {\bf 152}, 391 (1985); M. Meneguzzi {\it et al.},
{\it ibid} {\bf 182}, 169 (1987). The aspect ratio used by these authors
is $\Gamma = 1$ for the first and $\Gamma = 2 \sqrt{2}$ for the second.
\end{thebibliography}
\end{document}